\documentclass[amssymb, preprintnumbers, showpacs, showkeys, aps, prb,
               superscriptaddress, twocolumn, longbibliography]{revtex4-2}
\usepackage{color} 
\usepackage{xcolor} 
\usepackage[colorlinks,bookmarks=false,citecolor=blue,linkcolor=blue,urlcolor=blue]{hyperref}
\usepackage{slashed}
\usepackage{graphicx}
\usepackage{amsmath}
\usepackage{latexsym}
\usepackage{epstopdf}
\usepackage{amsmath}
\usepackage{enumitem}
\usepackage{amssymb,amsmath}
\usepackage{multirow}
\usepackage{mathtools}
\usepackage{bbm}
\usepackage{bm}
\usepackage{amsfonts}
\usepackage{amssymb}
\usepackage{calligra}
\usepackage{calrsfs}
\usepackage{makecell}
\usepackage{comment}
\usepackage[normalem]{ulem}
\usepackage[colorlinks, bookmarks=false,
            citecolor=blue, linkcolor=blue, urlcolor=blue]{hyperref}
\usepackage[USenglish,american]{babel}
\usepackage{subfigure}

\expandafter\ifx\csname package@font\endcsname\relax\else
 \expandafter\expandafter
 \expandafter\usepackage
 \expandafter\expandafter
 \expandafter{\csname package@font\endcsname}%
\fi
\begin{document}

\title{Quantum Phases and Structural Transitions of Bosons in Bubble-Shaped Confinements}

\author{Fabio Cinti}
\email{fabio.cinti@unifi.it}
\affiliation{Dipartimento di Fisica e Astronomia, Universit\`a di Firenze, I-50019, Sesto Fiorentino (FI), Italy}
\affiliation{INFN, Sezione di Firenze, I-50019, Sesto Fiorentino (FI), Italy}
         
\author{Gabriele Costa}
\email{gabriele.costa@studenti.unime.it}
\affiliation{Dipartimento di Scienze Matematiche e Informatiche, Scienze Fisiche e Scienze della Terra, Università degli Studi di Messina, Viale F. Stagno d’Alcontres 31, 98166 Messina, Italy}
\affiliation{Institute for Chemical-Physical Processes, National Research Council of Italy (IPCF-CNR),
98158 Messina (Italy)}

\author{Giuseppe Pellicane}
\email{giuseppe.pellicane@unime.it}
\affiliation{Dipartimento di Scienze Biomediche, Odontoiatriche e delle Immagini Morfologiche e Funzionali, Universit\`a degli Studi di Messina, I-98125, Messina, Italy}
\affiliation{School of Chemistry and Physics, University of Kwazulu-Natal, 3209 Pietermaritzburg, South Africa}
\affiliation{National Institute of Theoretical and Computational Sciences (NIThECS), 3209 Pietermaritzburg, South Africa}

\author{Santi Prestipino}
\email{sprestipino@unime.it}
\affiliation{Dipartimento di Scienze Matematiche e Informatiche, Scienze Fisiche e Scienze della Terra, Università degli Studi di Messina, Viale F. Stagno d’Alcontres 31, 98166 Messina, Italy}
\affiliation{Institute for Chemical-Physical Processes, National Research Council of Italy (IPCF-CNR),
98158 Messina (Italy)}
      
\author{Matteo Ciardi}
\email{matteo.ciardi@tuwien.ac.at}
\affiliation{Institute for Theoretical Physics, TU Wien, Wiedner Hauptstraße 8-10/136, 1040 Vienna, Austria}

\begin{abstract}
We investigate the structural properties and superfluid behavior of 
soft-core bosons confined in a spherical bubble trap at large particle 
numbers ($N \le 1800$). Using Path Integral Monte Carlo simulations with worm algorithm sampling in continuous space, we show that 
increasing $N$ drives a sequence of layering transitions: beyond the 
established bilayer supersolid, a third concentric layer of clusters emerges at $N \approx 1800$. By analyzing the distribution of permutation cycles 
distribution, we demonstrate that a finite superfluid fraction ($f_s \approx 10\%$) persists even in this multi-layer regime. 
These findings clarify the role of particle number in driving layer 
formation on curved manifolds, offering a concrete guide for future 
experiments with bubble-trapped bosons.
\end{abstract}

\date{\today}
\maketitle

\section{Introduction}\label{sec1}
Supersolidity, the simultaneous appearance of crystalline order and
superfluidity in the same many-body state \cite{boninsegni2012a}, has long been sought
in solid $^4$He \cite{kim2012} and ultracold bosonic gases \cite{cinti2010}. 
Bosons interacting through a soft-core potential, which is zero beyond a certain range and remains finite at zero separation, provide one of
the cleanest theoretical routes to capture the main features of the supersolid phase \cite{saccani2011}.
Above a critical density, the ground state spontaneously breaks translational symmetry to form a
cluster crystal, with each site occupied by a mesoscopic,
delocalized boson cluster, while the exchange of particles between
clusters sustains a finite global superfluid fraction~\cite{boninsegni2012a,henkel2010}.

It is well known that clusters are not 
specific to cold-atom systems and quantum systems \cite{Kalman2010}:
to some extent, clusterization occurs whenever the Fourier transform of the interparticle 
potential develops a negative region at some finite wavevector $q^\ast$, 
favoring a periodic density modulation of wavelength $2\pi/q^\ast$. This 
mechanism is shared by a broad class of soft-matter systems, from
colloidal suspensions to polymer-grafted nanoparticles~\cite{likos2001}. 
In ultracold atoms, cluster formation had initially been proposed through Rydberg-dressed interactions~\cite{weckesser2024science}.
Over the last ten years, experiments on lanthanide condensates, characterized by strong magnetic dipole moments, 
have emerged as the main platform to study there phenomena~\cite{bottcher2019,chomaz2019,tanzi2019,recati2023}. Finally, in a different context, bubble phases were investigated in two-dimensional electron systems ~\cite{ETTOUHAMI2004}. 

Some of us and coworkers have previously exploited this mechanism to observe supersolid cluster crystals in the presence of isolated defects~\cite{cinti2014natcomm} and,
in a closely related dipolar setting, self-bound filament structures in
free space~\cite{kora2019}. Most studies of this mechanism have been
carried out in flat space, either in harmonic traps  \cite{PhysRevA.86.013619} or in homogeneous systems with periodic boundary conditions \cite{saccani2012}. Curved geometries introduce a further, largely
unexplored ingredient: on a curved manifold, the interplay between
geometry and quantum correlations becomes directly observable, with no
counterpart in a standard planar or harmonic trap.

\begin{figure}[t]
\centering
\includegraphics[width=1\linewidth]{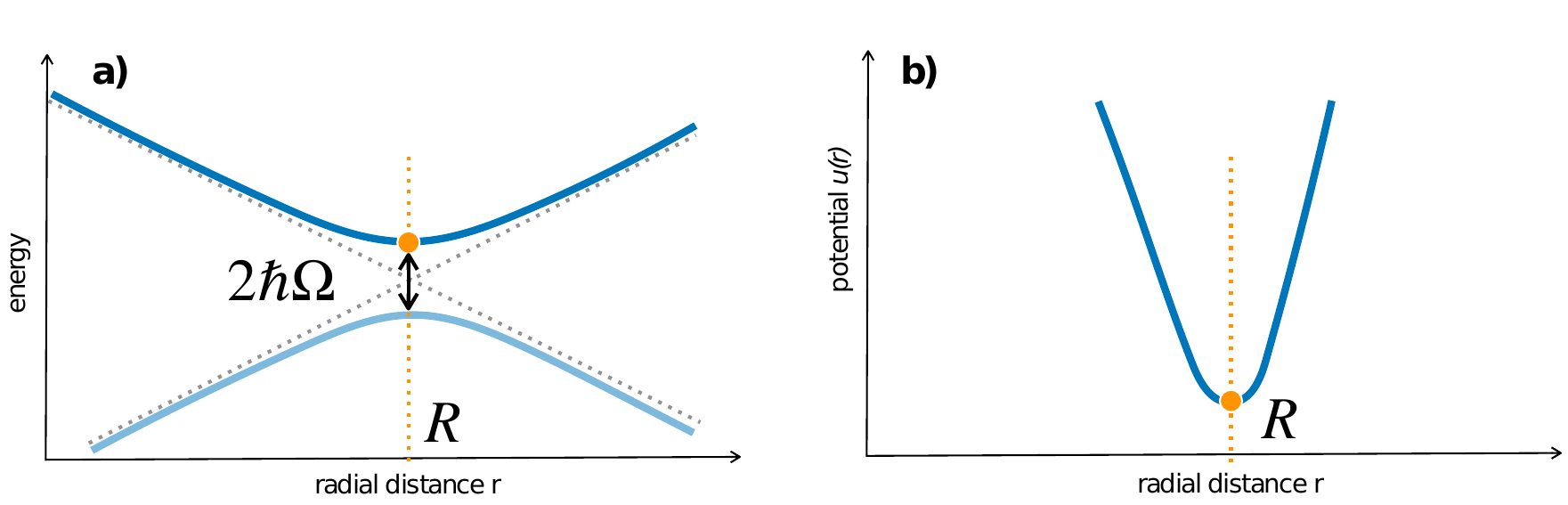}
\caption{Conceptual sketch of the bubble-trap mechanism. (a) The
radio-frequency dressing of the bare magnetic quadrupole opens an
avoided crossing of size $2\hbar\Omega$; atoms adiabatically follow the
populated upper branch. (b) The resulting effective radial potential
$u_{\rm ext}(r)$, Eq.~\eqref{eq:uext}, with its minimum at $r=R$.}
\label{fig:mechanism}
\end{figure}

A spherical shell is a particularly natural curved geometry to realize
experimentally, most often as the ``bubble'' trap. Physically, such a trap is realized starting from
a standard magnetic quadrupole trap. A radio-frequency (RF)
field, tuned close to resonance with the local Zeeman splitting, is then applied to couple
the trapped and untrapped spin sublevels. In the dressed-atom picture, the RF coupling opens an avoided crossing of size $2\hbar\Omega$:
Atoms adiabatically following the populated upper branch of this crossing then experience a
smooth three-dimensional potential with a minimum on a sphere of
radius $R$, replacing the original field zero (Fig.~\ref{fig:mechanism}a)
and yielding the bubble trap~\cite{zobay2001,zobay2004,garraway2016,veyron2026}.

\begin{figure}[t]
\centering
\includegraphics[width=0.8\linewidth,height=5cm,keepaspectratio]{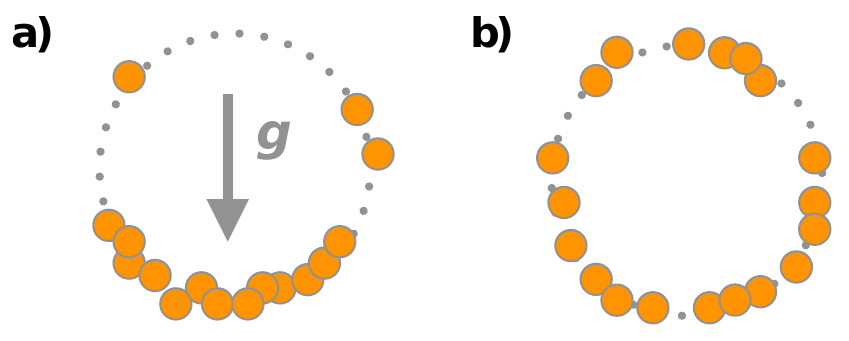}
\caption{(a) On Earth,
gravity is comparable to the weak curvature of the trap and the atomic
shell sags and collapses at the bottom. (b) In microgravity, the density remains uniformly distributed in the shell.}
\label{fig:gravity}
\end{figure}

On Earth, gravity is far from negligible compared to the weak curvature of this
potential, so the shell sags and collapses toward the bottom, breaking the
spherical symmetry (Fig.~\ref{fig:gravity}a); a genuinely uniform, closed
atomic shell has so far only been realized in microgravity, in
orbiting platforms~\cite{aveline2020,carollo2022,lundblad2023}
(Fig.~\ref{fig:gravity}b). This experimental difficulty has not stopped
theoretical progress: theory and numerical simulation remain the
primary source of quantitative predictions for these systems in the absence of a straightforward experimental realization.

In the last few years, extensive theoretical studies have explored the behavior of weakly interacting bosons on a sphere. Bereta \textit{et al.} obtained the semiclassical
critical temperature for Bose--Einstein condensation in spherical box and
shell traps~\cite{bereta2019}. Tononi and Salasnich showed that, on a zero-thickness two-dimensional sphere, curvature restores a finite
condensation temperature that would be forbidden in flat two-dimensional
space (corresponding to the limit of infinite radius)~\cite{tononisalasnich2019prl}. Quantum
Monte Carlo simulations have shown that, in a finite-thickness shell, superfluidity survives at the mesoscopic scale, up to a thousand
particles~\cite{tononicintisalasnich2020prl}; a subsequent review placed
these results in the broader context of shell-shaped atomic gases,
emphasizing that condensation and superfluidity are, in principle,
distinct transitions whose separation is itself controlled by
curvature~\cite{tononisalasnich2024review}. These studies, however,
are restricted to the weakly interacting, dilute regime. The natural
question is what happens when interactions are strong enough to drive the
system toward crystallization — the regime addressed by the present work.
Some clues come from mean-field theory at zero temperature~\cite{prestipino2019a}: at high density, soft-core bosons would organize themselves in clusters placed at the vertices of a regular or semiregular polyhedron.

Using Path Integral Monte Carlo (PIMC) simulations, we have characterized the bubble-trapped system in a sequence of three studies. In the first work \cite{ciardi2024prl}, we tuned the strength of quantum fluctuations (i.e., the boson mass) to drive a direct crystal-to-superfluid melting of a single layer of clusters, in good quantitative agreement with the corresponding two-dimensional flat-space result \cite{macri2013}. This picture was also discussed in Ref.~\cite{ciardi2025}, further including Earth's gravity effects in a small system. Extending the particle number to a few hundred, and simultaneously weakening the trapping potential, revealed a richer phase sequence: a cluster crystal state with polyhedral symmetry, for small quantum fluctuations; a bilayer supersolid, in which a second layer of clusters emerges while crystalline order persists, for intermediate fluctuations; and a uniform, structureless superfluid shell, at larger fluctuations.
The present paper extends this picture pushing the PIMC simulations to a previously unexplored large-$N$ regime, where an additional, third layer appears.

The paper is organized as follows. In Section~\ref{sec2}, we introduce the model and the simulation method. In Section~\ref{sec3}, we describe the phase sequence summarized above at relatively small $N$. In Section~\ref{sec4}, we present the large-$N$ results. Finally, in Section~\ref{sec5} we concludes and give some outlooks.

\section{Model and Method}\label{sec2}

\subsection{Soft-Core Bosons in a Spherical Shell}\label{sec2a}

\begin{figure}[t]
\centering
\includegraphics[width=0.8\linewidth,height=5cm,keepaspectratio]{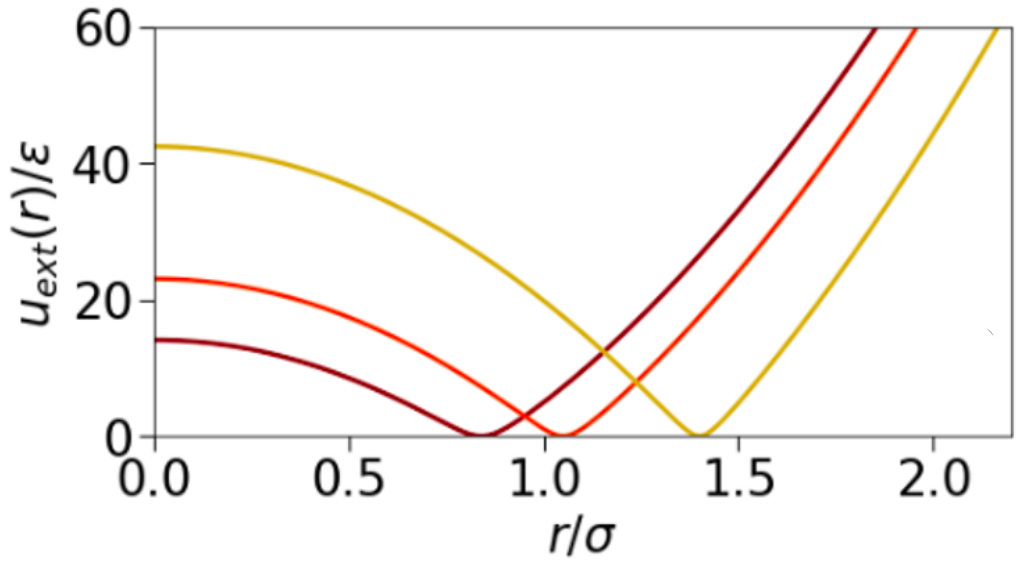}
\caption{Bubble-trap potential $u_{\rm ext}(r)$, Eq.~\eqref{eq:uext}, for $\Delta$=0.70 (brown), 1.10 (orange), and 1.96 (gold), and $\Omega=0.0441$.}
\label{fig:thinshell}
\end{figure}

We consider a system of $N$ spinless bosons of mass $m$, 
confined in a spherical
shell of radius $R$ by a weak bubble-trap potential. The system is
described by the Hamiltonian
\begin{equation}
H = \sum_{i=1}^{N}\left[-\lambda\nabla_i^2 + u_{\rm ext}(|\mathbf{r}_i|)\right]
+ \sum_{i<j} u_{\rm int}(|\mathbf{r}_i-\mathbf{r}_j|) \,,
\label{eq:hamiltonian}
\end{equation}
with $\lambda=\hbar^2/(2m)$.
The two-body interaction, $u_{\rm int}(r)$, is a soft, purely repulsive square barrier of height $\epsilon$ and width
$\sigma$, i.e., $u_{\rm int}(r)=\epsilon$ for $r<\sigma$ and zero
otherwise. Unlike a hard core, this potential remains finite at zero
separation, allowing multiple bosons to occupy the same
cluster at moderate energy cost, which represents the microscopic origin of
cluster-crystal behavior. The external confinement
$u_{\rm ext}(r)$ is the effective bubble-trap potential already
introduced in Section~\ref{sec1}. 
In our simulations, we use the standard parametrized form for the resulting 
effective potential,
following Sun \textit{et al.}~\cite{sun2018},

\begin{equation}
u_{\rm ext}(r) = \frac{u_0}{\Omega}\sqrt{\frac{(r^2-\Delta)^2}{4}+\Omega^2} - u_0\,,
\label{eq:uext}
\end{equation}
with $u_0$ the barrier height. $\Delta$ and $\Omega$ are square-length parameters related to the detuning and Rabi frequency of the rf field, respectively. In the thin-shell limit (i.e. $\Omega \ll u_0$) Eq.~\eqref{eq:uext} becomes harmonic around the minimum at $R=\sqrt{\Delta}$.
As a result, the confinement is tight
enough that only small radial displacements $x \equiv r-R$ away from the
shell are relevant, and Eq.~\eqref{eq:uext} can be Taylor-expanded around
its minimum. Writing $r^2-R^2 \simeq 2Rx$ and expanding the square root to
leading order in $x$ gives a simple radial harmonic well,
\(u_{\rm ext}(x) \simeq \frac{u_0 R^2}{2\Omega^2}\,x^2\).
Fig.~\ref{fig:thinshell} shows some examples of the bubble trap potential for some values of $\Delta$.

We use the soft-core parameters $\epsilon$ and $\sigma$ as our
energy and length units, and express temperatures in units of $\epsilon$
with $k_B=1$. The parameter $\lambda$ sets the strength
of quantum fluctuations relative to the interaction and confinement
energy scales. In accordance with Ref.~\cite{cinti2026pra}, in this paper
we take $R=1.15$, $u_0=1.0$, and $\Omega=0.0441$. 
Considering a number of bosons $N=200$, one obtains the icosahedral 
single-shell supersolid at $T=0.5$. The phase sequence recalled in Section~\ref{sec3} is, in turn, obtained for $N=600$ by varying $\lambda$, while keeping $R$, $u_0$, and $\Omega$ fixed. The results of Section~\ref{sec4} are obtained with the same set of 
parameters, but increasing $N$ up to 1800.

\subsection{Path Integral Monte Carlo}\label{sec2b}

We compute the finite-temperature equilibrium properties of the system,
at fixed particle number $N$ (canonical ensemble), by means of
continuous-space Path Integral Monte Carlo (PIMC)
simulations~\cite{ceperley1995}. Each boson is represented by an 
imaginary-time worldline, obtained by discretizing the inverse
temperature $\beta = 1/T$ into $M$ time slices. Particle exchanges,
essential to sample Bose statistics, are handled through the
continuous-space worm algorithm~\cite{boninsegni2006}, 
which allows worldlines to open into a
disconnected $worm$, and efficiently reconnect them into permutation
cycles of arbitrary length. 

Here, we are going to use two indicators of the transitions: 
the particle density and the superfluid response. 
Regarding the first indicator, it is convenient to work with 
the integrated radial density
\begin{equation}
n(r) = r^2 \int d\theta\, d\varphi\, \sin\theta\, \rho(r,\theta,\varphi),
~ N = \int dr\, n(r)\,.
\label{eq:nr}
\end{equation}
$\rho(r,\theta,\varphi)$ is the density in spherical coordinates; 
$n(r)$ gives the number of particles in an infinitesimal spherical shell at a distance $r$ from the trap center. This quantity will be useful for discussing some structural properties in Secs.~3 and~4.

The superfluid response is obtained through the area
estimator~\cite{sindzingreklein1989}, generalized to a spatially
resolved (``zonal'') form following Ref.~\cite{ciardi2022entropy}. 
Considering a three-dimensional system, this estimator yields: 
\begin{equation}\label{eq:azlocal}
	\textbf{A}(\textbf{r}) =  \frac{1}{2} \sum_{i=1}^{N} \sum_{m=0}^{M-1} \textbf{r} \times \textbf{r}_{i}^{m+1} \delta^3(\textbf{r}-\textbf{r}_{i}^{m}) \,, 
\end{equation}
where $\mathbf{r}_i^{\,m}$ is the bead of particle $i$ on time slice $m$, and 
$\textbf{A}(\textbf{r})=(A_x(\mathbf{r}),A_y(\mathbf{r}),A_z(\mathbf{r}))$, each component representing the worldline contribution to the 
projected area on the corresponding  perpendicular plane. Now choosing one of the three axis (say $z$), the superfluid density $\rho^{(z)}_s(\mathbf{r})$ reads
\begin{equation}
\rho^{(z)}_s(\mathbf{r}) = \frac{4m^2\left[\langle A_z A_z(\mathbf{r})\rangle
- \langle A_z\rangle\langle A_z(\mathbf{r})\rangle\right]}
{\hbar^2\beta\, r_\perp^2}\,,
\label{eq:rhos}
\end{equation}
$r_\perp = \sqrt{x^2+y^2}$, and $\rho^{(x)}_s(\mathbf{r})$, $\rho^{(y)}_s(\mathbf{r})$ likewise. In this way one has 
the superfluid fraction in a given region $k$ 
\begin{equation}
\label{pippo}
f_{s,k}^{(i)} = \frac{1}{I_{cl,k}^{(i)}} \int_k d\textbf{r} \; \rho_s^{(i)}(\textbf{r}) r_\perp^2 \, ,
\end{equation}
where $I^{(i)}_{cl,k}$ is the classical moment of inertia, measured around the $i$-axis, of particles in region $k$ \cite{cinti2026pra}. 
It seems appropriate to also use the integrated superfluid radial density, which reads
\begin{equation}
\label{pippo1}
n_s^{(i)}(r) = r^2 \int{\rm d}\theta\,{\rm d}\phi \, \sin \theta \, \rho_{s}^{(i)}(r) \,. 
\end{equation}
We note that Eq.~\eqref{pippo1} is useful to be compared with  $n(r)$ in Eq.~\eqref{eq:nr}.

Finally, we characterize superfluidity directly through the statistics
of the permutation cycles, again sampled by using the worm algorithm. A given
worldline configuration decomposes into a set of closed permutation cycles $\mathcal{C}$, each involving $\ell(c)$ particles. In this way, one can define a histogram, $P$, which represents the distribution of cycle lengths:
\begin{equation}
P(L)=\frac{L}{N}\left\langle\sum_{c\,\in\,\mathcal{C}}\delta_{L,\,\ell(c)}\right\rangle\,
\label{eq:permutation}
\end{equation}
normalized so that $\sum_L P(L)=1$, i.e., $P(L)$ is the probability that a
randomly chosen worldline (particle) belongs to a permutation cycle of length $L$.
A system with only short cycles ($\text{max} \{ L \}  \approx 1$) behaves as an assembly of
distinguishable particles, while a sizable probability of long cycles,
$\text{max} \{ L \} \approx N$, signals extensive particle exchange and is a
qualitative signature of superfluidity.

\begin{figure}[t]
\centering
\includegraphics[width=1\linewidth]{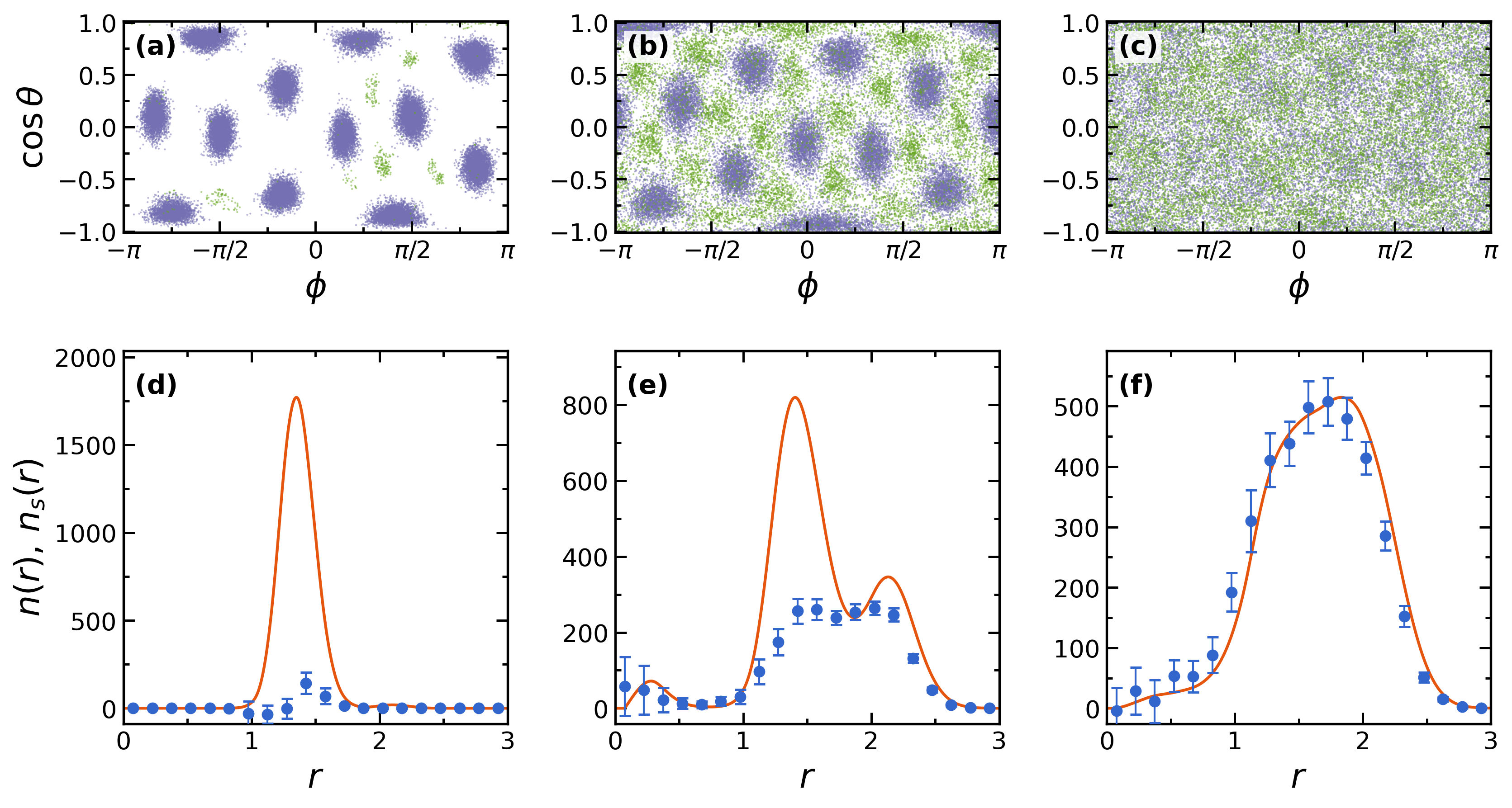}
\caption{Structural transition with increasing quantum fluctuations
$\lambda$, for $N=600$, $R=1.15$, and $T=0.5$. (a)--(c) Worldline
configurations projected onto the $(\varphi,\cos\theta)$ plane, colored by
shell (purple: inner shell; green: outer shell), from a single-shell
cluster crystal ($\lambda=0.10$) through a bilayer supersolid
($\lambda=0.35$) to a uniform superfluid shell ($\lambda=0.40$).
(d)--(f) Corresponding integrated radial density $n(r)$ (red) and
superfluid radial density $n_s(r)$ (blue circles), showing the growth 
of the superfluid component between and
eventually across the shells.}
\label{fig:centralresult}
\end{figure}

\section{Superfluidity of soft-core bosons in shallow spherical traps}\label{sec3}

The crystal-supersolid-superfluid transition driven by increasing
quantum fluctuations $\lambda$ was first established in the
single-layer setting of Ref.~\cite{ciardi2024prl} and was found to be
robust against gravity in Ref.~\cite{ciardi2025}. Here we summarize
the two-layer generalization of this behavior. In particular,
Fig.~\ref{fig:centralresult} shows the three regimes in two rows
of panels. The top row, Fig.~\ref{fig:centralresult}(a)–(c), 
displays snapshots of the worldline configurations projected onto the $(\varphi,\cos\theta)$ plane, where $\theta$ is the polar angle 
and $\phi$ is the azimuthal angle.
The bottom row, Fig.~\ref{fig:centralresult}(d)-(f), outlines 
the corresponding integrated densities $n(r)$, Eq.~\eqref{eq:nr}, 
and $n_s(r)$, i.e. Eq.~\eqref{pippo1} averaged over $x$, $y$, and $z$-axis. 
Panels (a) and (d) of Fig.~\ref{fig:centralresult} are obtained considering 
$\lambda=0.10$. In this case we observe a one-layer cluster crystal.
By increasing $\lambda$ to 0.35, see panels (b) and (e), the system clearly
displays a bilayer structure which indeed turns out to be supersolid. The third phase -- panels (c) and (f) -- gives a uniform superfluid shell ($\lambda=0.40$).

Fig.~\ref{fig:centralresult} summarizes a clear behavior: as $\lambda$  increases, the
single-layer crystal splits into two concentric layers,
visible as the two peaks of $n(r)$ in panel (e). Remarkably, the developed supersolidity is not confined within each layer individually,
but it bridges the inner and outer layers; indeed $n_s(r)$
spans both peaks (see again panel (e)). It is also useful here to report and discuss the global superfluid fraction, $f_s(\lambda)$. This quantity can be, for instance, obtained by summing the zonal contributions $f^{(i)}_{s,k}$, Eq.~\eqref{pippo}, over all regions $k$ of the system. The symmetry of the potential ensures the same value of $f^{(i)}_{s}$ for all principal axes ($i=x,\,y,\,\text{and}\,z$) within statistical errors. So, in the single-layer crystal phase we have
$f_s(\lambda=0.10)=0.028(2)$. Such a small value refers to a superfluidity of bosons that remains localized into droplets. Instead, the supersolid phase has $f_s(\lambda=0.30)=0.64(2)$. Finally, $f_s(\lambda=0.40)=1.01(4)$ as expected looking at Fig.~\ref{fig:centralresult}(c) and (f).

The polyhedral symmetry of the clusters, already apparent in the
single-layer crystal, persists in the bilayer supersolid. Fixing for
instance $\lambda=0.16$ (i.e., at the edge between the one-layer
crystal and the bilayer supersolid), the two nested layers sit at the
vertices of an icosahedron ($12$ clusters, inner layer) and of the dual
dodecahedron ($20$ clusters, outer layer), perfectly interlocked into a
pentakis dodecahedron~\cite{cinti2026pra}.

Unlike soft-core potential models, magnetic dipolar atoms like 
dysprosium offer a real physical platform to realize quantum cluster phases.
For this reason, in Ref.~\cite{ciardi2024prl} we also
explored dipolar bosons confined in thin-shell bubble traps. 
By increasing the $s$-wave scattering length $a_s$, 
we observed that the system exhibits a crystal-supersolid-superfluid 
progression qualitatively similar to the one discussed earlier. 
In particular, the main difference lies
in the presence of a ring of well-defined clusters
around the equator in the crystal regime. 
Subsequently, for higher values of $a_s$, 
a modulated ``equatorial supersolid'' is obtained,
and finally, the atom aggregates dissolve into a uniform superfluid ribbon.
The anisotropy of the dipolar interaction produces a different
superfluid response along the polarization axis, $z$, compared to the
transverse directions $x$, $y$, an effect entirely absent for the
isotropic soft-core potential in Eq.~\eqref{eq:hamiltonian}.

We now turn to systems with larger $N$, where the layering sequence is
pushed beyond the bilayer.

\section{Multi-layers in Bubble Traps}\label{sec4}

\begin{figure}[t]
\centering
\includegraphics[width=1\linewidth]{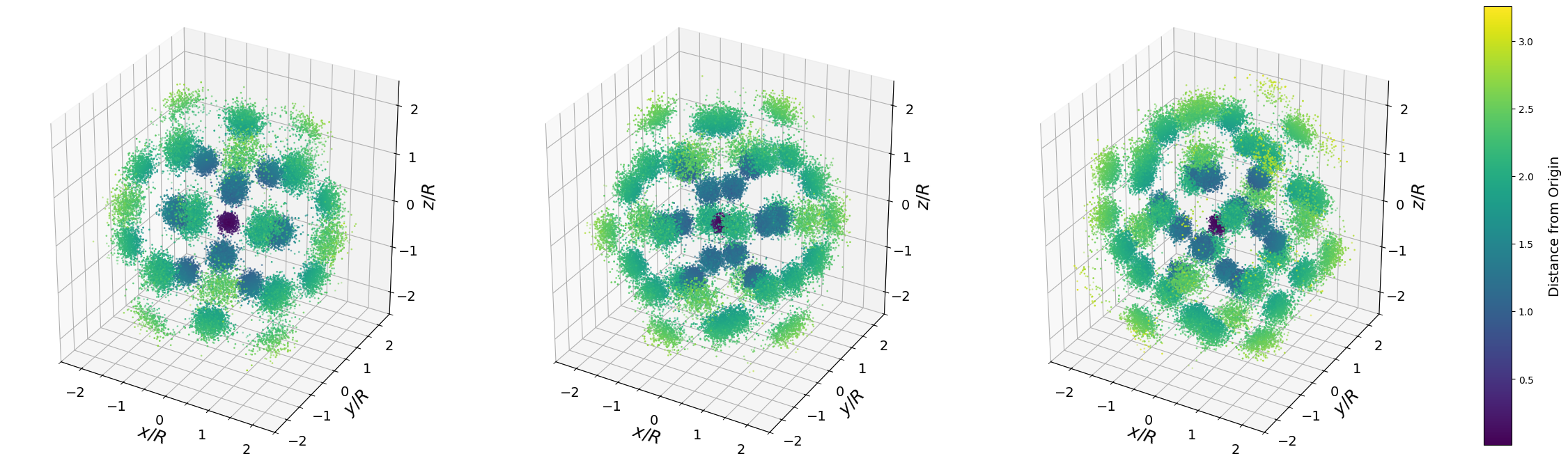} 
\caption{ Worldline snapshots for $N=1200$, $1500$ and $1800$
(left to right), colored by distance from the trap center. A third,
outer shell of clusters (green-yellow) nucleates beyond the two shells
of the bilayer supersolid (dark purple and blue-green) as $N$ is
increased.}
\label{fig:largeNthirdlayer}
\end{figure}

We extend the study of Section~\ref{sec3} to larger values of $N$, performing PIMC simulations at $N=1200$, $1500$, and $1800$, again keeping $R=1.15$, $u_0=1$, $\Omega=0.0441$, $T=0.5$, and $\lambda=0.16$. The worldline snapshots of the sampled configurations are shown in Fig.~\ref{fig:largeNthirdlayer}, where the worldline beads are colored by their radial distance from the trap center. Three concentric families of clusters are visible: an inner, nearly undistorted cluster close to the origin (dark purple), an intermediate layer of well-separated clusters (blue-green), and an outer, more diffuse layer (green-yellow) whose clusters are less sharply localized than those of the inner two layers and become more pronounced as $N$ increases.

\begin{figure}[t]
\centering
\includegraphics[width=1\linewidth,keepaspectratio]{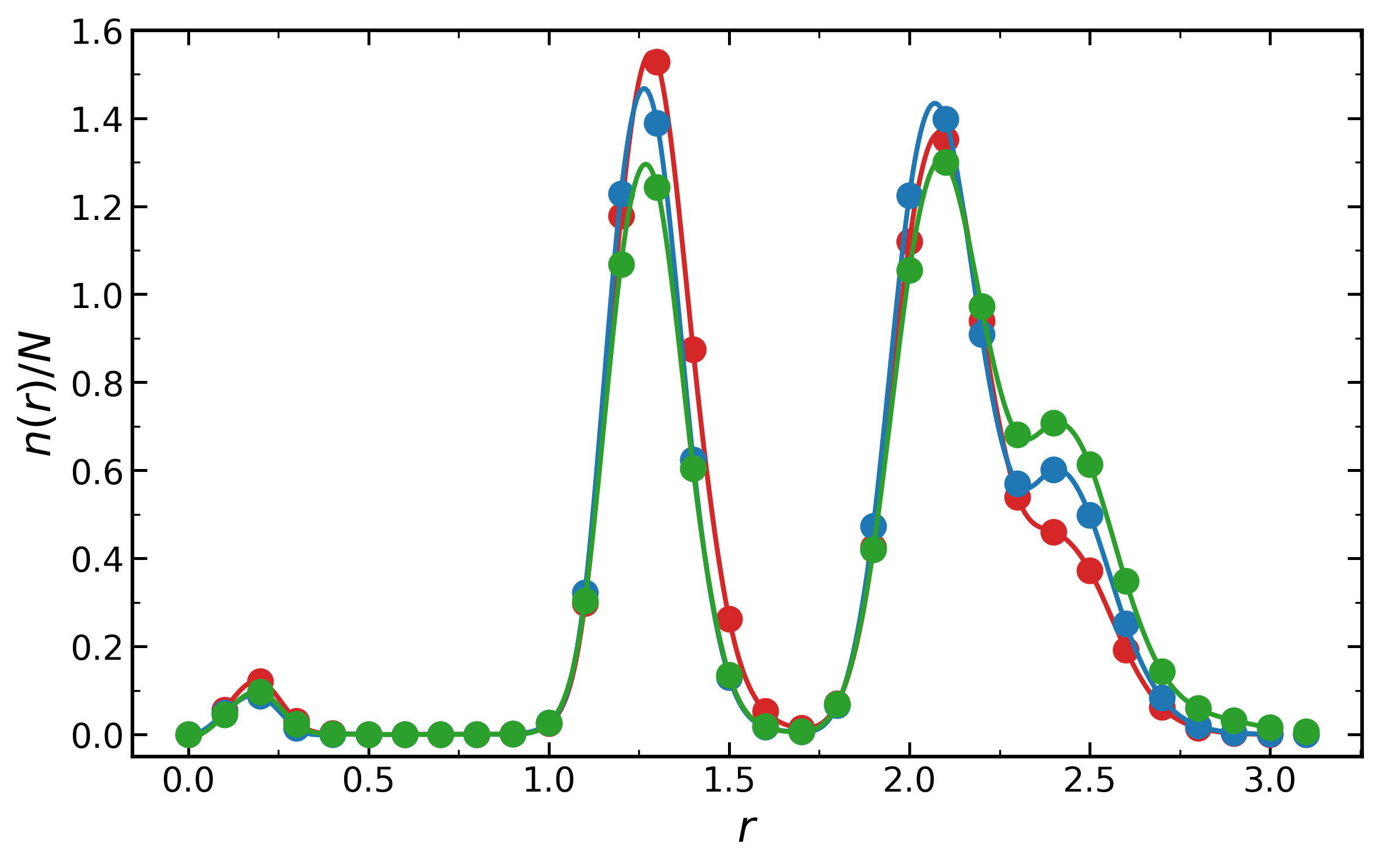}
\caption{Integrated radial density $n(r)/N$, Eq.~\eqref{eq:nr}, for $N=1200$ (red), $1500$ (blue), and $1800$ (green) bosons at $R=1.15$, $u_0=1$, $\Omega=0.0441$, $T=0.5$, and $\lambda=0.16$. Beyond the two peaks already present in the bilayer regime
(Fig.~\ref{fig:centralresult}), a third, outer shoulder grows into a
distinct peak as $N$ increases, consistent with the snapshot configurations of Fig.~\ref{fig:largeNthirdlayer}.}
\label{fig:largeNnr}
\end{figure}

This third family of clusters is confirmed quantitatively by the
integrated radial density $n(r)/N$, Eq.~\eqref{eq:nr}, shown in
Fig.~\ref{fig:largeNnr} for the three system sizes. In addition to the
two peaks already identified, a third, broader
feature develops at larger $r$ as $N$ increases from $1200$ to $1800$.
The figure shows a shoulder barely visible at $N=1200$ that
grows into a distinct peak at $N=1800$, consistent with 
Fig.~\ref{fig:largeNthirdlayer} and signaling the emergence of a third 
layer of clusters beyond the outer layer.

\begin{figure}[t]
\centering
\includegraphics[width=1\linewidth]{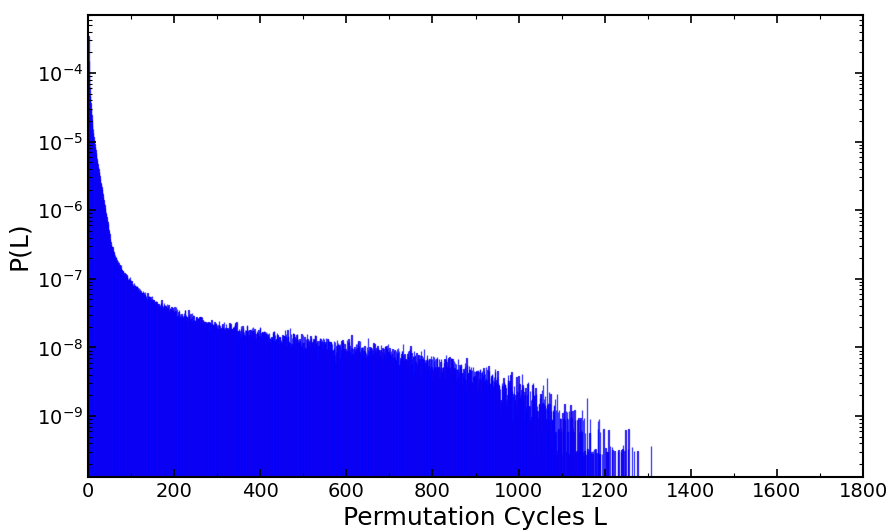}
\caption{Permutation cycle-length distribution $P(L)$ for $N=1800$
bosons, at the same parameters as Fig.~\ref{fig:largeNnr}.}
\label{fig:permutations}
\end{figure}

We characterize the quantum features in this large-$N$ regime by analyzing 
the cycle-length distribution $P(L)$ of Eq.~\eqref{eq:permutation}. 
Figure~\ref{fig:permutations} shows $P(L)$ for $N=1800$: the distribution 
develops an extended tail of long permutation cycles that persists up to 
$L\approx 1100$. This tail is the direct microscopic signature of superfluid 
exchange discussed in Section~\ref{sec2b}. Specifically, it indicates that, even 
for such a large $N$, a non-negligible number of particles participate in 
long permutation cycles rather than remaining localized. For this system 
($T=0.5$), we estimated a finite superfluid fraction of $f_s\approx10\%$, 
confirming that superfluidity survives in the multi-layer regime at large-$N$.

\section{Conclusions and Outlook}\label{sec5}

The results summarized in this work confirm and extend the picture
established in Ref.~\cite{cinti2026pra}: a weak, spherically symmetric
bubble trap can hold several concentric layers of soft-core boson
clusters in place, with a sizable superfluid signal. The
large-$N$ PIMC simulations of Section~\ref{sec4} show that a third concentric
layer starts to be defined for $N\approx1800$. 
We believe that the present findings open the door to exploring a broader
range of curved geometries. Experimentally, 
realizing this system would require loading atoms with soft-core-like
interactions into a bubble trap, which, to our knowledge, remains
beyond current capabilities. On the other hand, dipolar species such as
those discussed in Section~\ref{sec3} may offer a more immediate, even though
qualitatively different, testbed~\cite{lundblad2019npj}. Should a
suitable platform become available, rotating potential barriers could
be used to probe dynamical signatures of the supersolid
phase, see e.g. Ref.~\cite{nesti2026pra}. 
Finally, further investigations based on a lattice-gas model, 
derived as a limiting case of an extended Bose-Hubbard model on the 
grid set by the cluster centers~\cite{prestipino2021pra}, could pave the way toward a general theoretical framework for layered supersolids in curved geometries.

\begin{acknowledgments}
This work was supported by the European Union through the Next Generation EU funds through the Italian MUR National Recovery and Resilience Plan, Mission 4 Component 2 - Investment 1.4 - National Center for HPC, Big Data and Quantum Computing (CUP B83C22002830001). F.~C. acknowledges financial support from PNRR MUR Project No. PE0000023-NQSTI. M.~C. acknowledges funding from  the Austrian Science Fund (FWF) through the cluster of excellence quantA (Grant No.\ 10.55776/COE1), and from the European Union’s Horizon Europe research and innovation program under the Marie Sklodowska-Curie Grant Agreement No. 101205948 (Q-UDiM). F.~C. and G.~P. acknowledge the NICIS Centre for High Perfomance Computing, South Africa, for providing computational resources.
\end{acknowledgments}


%

\end{document}